\documentclass[sigconf]{acmart}

\AtBeginDocument{%
  \providecommand\BibTeX{{%
    \normalfont B\kern-0.5em{\scshape i\kern-0.25em b}\kern-0.8em\TeX}}}

\copyrightyear{2022} 
\acmYear{2022} 
\setcopyright{acmcopyright}\acmConference[CIKM '22]{Proceedings of the 31th ACM International Conference on Information and Knowledge Management}{November 1--5, 2021}{Queensland, Australia}
\acmBooktitle{Proceedings of the 30th ACM International Conference on Information and Knowledge Management (CIKM '22), November 1--5, 2021, Queensland, Australia}
\acmPrice{15.00}
\acmDOI{10.1145/3340531.3411927}
\acmISBN{978-1-4503-6859-9/20/10}

\renewcommand\footnotetextcopyrightpermission[1]{} 

\usepackage{diagbox}

\usepackage{float}
\usepackage{array}
\usepackage{multirow}
\usepackage{threeparttable}
\usepackage{subfigure}
\usepackage{bbding}
\usepackage{enumitem}
\setlist[itemize]{leftmargin=*}

\usepackage[linesnumbered,ruled]{algorithm2e}

\begin{document}
\fancyhead{}
\title{Coarse-to-Fine Knowledge-Enhanced Multi-Interest Learning Framework for Multi-Behavior Recommendation}

\author{Chang Meng$^{\ast 1}$, Ziqi Zhao$^{2}$, Wei Guo$^{3}$, Yingxue Zhang$^{4}$, \\ Haolun Wu$^{5}$, Chen Gao$^{6}$, Dong Li$^{3}$, Xiu Li$^{1}$, Ruiming Tang$^{3}$}

\affiliation{$^{1}$Shenzhen International Graduate School, Tsinghua University~~~$^{2}$Harbin Engineering University \\
$^{3}$Noah's Ark Lab, Huawei $^{4}$Huawei Technologies Canada\\
$^{5}$McGill University $^{6}$Department of Electronic Engineering, Tsinghua University
\\
mengc21@mails.tsinghua.edu.cn, zziqi@hrbeu.edu.cn\\
\{guowei67, yingxue.zhang, lidong106, tangruiming\}@huawei.com\\
haolun.wu@mail.mcgill.ca, chgao96@gmail.com, li.xiu@sz.tsinghua.edu.cn}

\renewcommand{\shortauthors}{}

\begin{abstract}

Multi-types of behaviors (e.g., clicking, adding to cart, purchasing, etc.) widely exist in most real-world recommendation scenarios, which are beneficial to learn users’ multi-faceted preferences. 
As dependencies are explicitly exhibited by the multiple types of behaviors, effectively modeling complex behavior dependencies is crucial for multi-behavior prediction.
The state-of-the-art multi-behavior models learn behavior dependencies indistinguishably with all historical interactions as input.
However, different behaviors may reflect different aspects of user preference, which means that some irrelevant interactions may play as noises to the target behavior to be predicted.
To address the aforementioned limitations, we introduce multi-interest learning to the multi-behavior recommendation.
More specifically, we propose a novel Coarse-to-fine Knowledge-enhanced Multi-interest Learning (CKML) framework to learn shared and behavior-specific interests for different behaviors.
CKML introduces two advanced modules, namely \emph{Coarse-grained Interest Extracting (CIE)} and \emph{Fine-grained Behavioral Correlation (FBC)}, which work jointly to capture fine-grained behavioral dependencies. CIE uses knowledge-aware information to extract initial representations of each interest. FBC incorporates a dynamic routing scheme to further assign each behavior among interests. Additionally, we use the self-attention mechanism to correlate different behavioral information at the interest level. 
Empirical results on three real-world datasets verify the effectiveness and efficiency of our model in exploiting multi-behavior data. Further experiments demonstrate the effectiveness of each module and the robustness and superiority of the shared and specific modelling paradigm for multi-behavior data.

\noindent\let\thefootnote\relax\footnotetext{$\ast$ Work done when he is a research intern at Noah’s Ark Lab, Huawei.}
\end{abstract}




\settopmatter{printacmref=false, printfolios=false}
\maketitle


\section{Introduction}
\label{intro}

\begin{figure}[t]
	\centering
	\setlength{\abovecaptionskip}{0cm}
	\setlength{\belowcaptionskip}{-5mm}
	\includegraphics[width=0.47\textwidth]{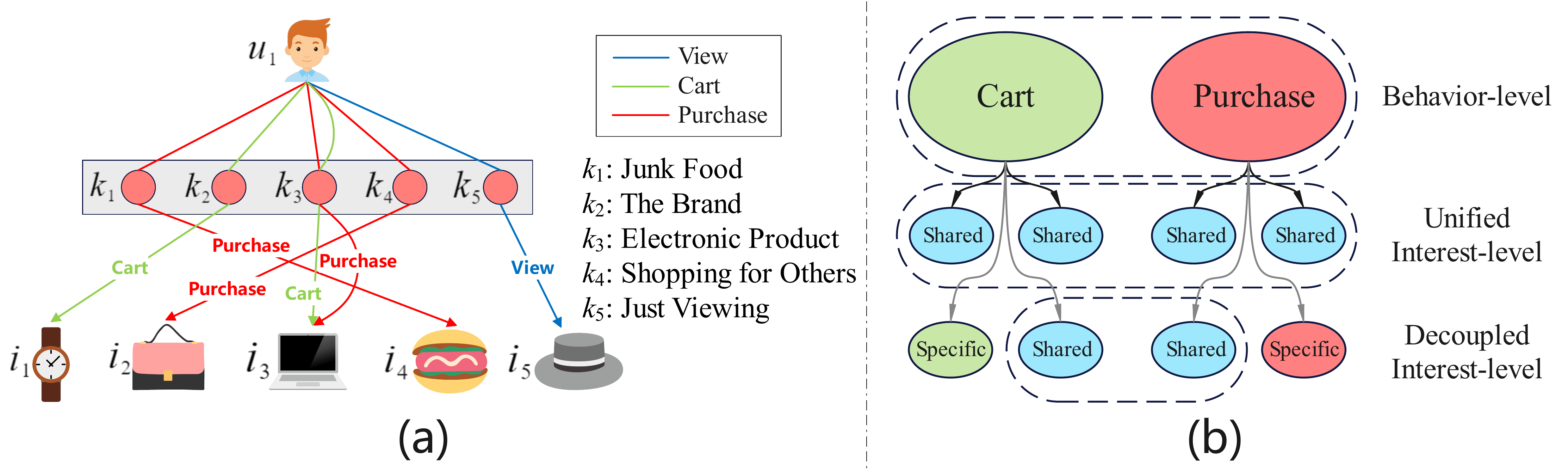}
	\caption{(a) An example of multi-faceted interests behind multiple behaviors on an e-commerce scenario. (b) An example of behavior correlation at different granularities. Black and gray arrows represent different interest divisions, and dashed boxes represent correlations of information.}
	
	\label{fig:relation}
	\vspace{1mm}
	
\end{figure}

Collaboration filtering (CF) \cite{cfsurvey} is widely used in industry to deeply probe into the latent information behind users’ behaviors. 
CF firstly learns representations for both users and items from their historical interactions and then leverages these representations to make predictions.
Most of the existing CF methods \cite{biasmf,dmf,gcn,ngcf,lightgcn} are designed to model a single behavior. 
However, users usually have interactions with items through various behaviors in real-world applications, like viewing, tagging as favorites, carting, and purchasing in the e-commerce scenario.
As various behaviors may express users' complementary interests with items, utilizing the multi-behavior data simultaneously is necessary.

Many research efforts have been devoted to this problem to better capture collaborative signals from multi-behavior data, which can be divided into two categories \cite{huang2021recent}.
The first category tackles the multi-behavior recommendation problem with advanced neural networks. 
For example, DIPN \cite{dipn} and MATN \cite{matn} apply the attention network and transformer network separately to model dependence between different behaviors. 
Further, MBGCN \cite{mbgcn} and KHGT \cite{khgt} strengthen multi-behavior data with graph neural network (GNN) \cite{kipf2016semi,gcn,gat} by regarding multiple behaviors data as heterogeneous graphs to learn more complex heterogeneous information behind the interaction between users and items.
The second category utilizes multi-behavior data with multi-task learning (MTL) \cite{caruana1997multitask}.
For instance, NMTR \cite{nmtr} leverages all behaviors of users as prediction targets to improve the learning of users and items.
EHCF \cite{chen2020efficient} and GHCF \cite{ghcf} then improve NMTR \cite{nmtr} with knowledge transfer and GNN to correlate the predictions, respectively. 


However, these multi-behavior recommendation methods ignore the multi-faceted interests behind different behaviors.
As shown in Figure \ref{fig:relation}(a), here we exhibit a toy example, a user ($u_{1}$) generally adopts different items ($i_{1},i_{2},...,i_{5}$) under different behaviors. 
Each behavior belongs to a plurality of different interests ($k_{1},k_{2},...,k_{5}$), and each interaction is based on one or more interests. 
We can observe that different interests under the same behaviors may have different effects on the prediction of target behaviors.
For example, interests $k_2$ and $k_3$ are both under the behavior of carting. However, for predicting whether the user will purchase $i_3$, interest $k_3$ is more effective and meaningful, because the user add $i_3$ to cart based on $k_3$ and add the less relevant item $i_1$ to cart based on $k_2$.
And we can also find that some interests are specific only to a certain behavior (e.g. user $u_1$ adds item $i_1$ to cart only because of interest $k_2$)  
, and these interests may be noises for prediction of other behaviors. 
Fine-grained decoupling of behavior to interest-level representations can make full use of the potential dependence information in a delicate way, thus achieves both better interpretability and possibly superior performance.
So it is vital to explore the relationships of multi-behaviors at a level of multi-interests.

Some recent works have attempted to leverage multi-interest learning for recommendation, either utilize behavioral collaborative signals to learn multi-interest representations, or leverage structured relational information to construct multi-interest representations.
For the approaches fall under the first category, MIND \cite{mind}, ComiRec \cite{comirec} and DGCF \cite{dgcf} implicitly cluster historical user interactions by using powerful encoders, like dynamic routing and self-attention.
For the latter category, KGIN \cite{kgin}
and KTUP \cite{ktup} seek to leverage the auxiliary semantic information of knowledge graph to model multi-interests.
Despite the effectiveness of these methods, they are all designed for single behavior, which share two common limitations when applied for multi-behavior recommendation:
\begin{itemize}
    \item \textbf{Suboptimal Correlation Modeling. 
    }
    Existing multi-interest methods are designed for single behavior, all of which adopts a group of unified interests for each user.
    However, as illustrated above, 
    different interests under the same behaviors may have different effects on the prediction of target behaviors, and some interests are specific only to one behavior. 
    If we unify all interests for each behavior to model behavior correlation, it may inevitably introduce noise, 
    as some behavior-specific interests will negatively affect the learning of interest representations under other behaviors.
    We named this correlation modeling strategy as \textsl{unified interest-level}, as shown in Figure \ref{fig:relation}  (b). The \textsl{unified interest-level} roughly correlate the divided shared interests of one behavior with the ones of other behaviors, which will lead to a suboptimal modeling of correlation.
    So, a more reasonable strategy is necessary to better model behavior correlations. 
    \item \textbf{Difficulties in Interest Learning.}
    The learning of interest can be regarded as the grouping of users' historical behaviors through \emph{clustering}, where items from one cluster are expected to be closely related and collectively represent a specific aspect of user interests \cite{mind}. 
    In clustering theory, the final results are sensitive to the \emph{initialization of clustering centers}, which has been claimed in lots of works such as K-Means++ \cite{kmeans++}, K-Means$\|$ \cite{K-Means||} and Canopy \cite{Canopy}.
    Existing multi-interest methods like MIND \cite{mind} and DGCF \cite{dgcf} initialize clustering centers with random vectors, which lead to sub-optimal results as the generated centers may be very close to each other.
    Besides the \emph{initialization of clustering centers}, the \emph{process of clustering} via collaborative signals are also important for the clustering results. 
    Methods like KGIN \cite{kgin} and KTUP \cite{ktup} utilize the semantic information from knowledge graph to learn interest representation.
    However, they are possibly sub-optimal either, as they overlooks the rich collaborative signals that can be used for interest representation learning.
    As a result, a more flexible method which can make the initial center of interest as far as possible by using semantic information obtained from knowledge-aware information and make full use of collaborative signals' information in clustering process is needed. 
\end{itemize}

To tackle these two limitations, we propose a \textbf{\underline{C}}oarse-to-fine \textbf{\underline{K}}nowledge-enhanced \textbf{\underline{M}}uti-interest \textbf{\underline{L}}earning framework (CKML).
To handle the first limitation, CKML decouples interests into shared and behavior-specific parts for each behavior, then model the behavior correlation under the \textsl{decoupled interest-level}, as shown in Figure \ref{fig:relation}(b).
To tackle the difficulties for interest learning process, CKML leverages a Coarse-to-fine strategy to initialize the interest centers and then allocates different interactions to different interests through the collaborative signals.
Concretely, CKML consists of two modules: \textbf{\underline{C}}oarse-grained \textbf{\underline{I}}nterest \textbf{\underline{E}}xtracting (CIE) module and \textbf{\underline{F}}ine-grained \textbf{\underline{B}}ehavioral \textbf{\underline{C}}orrelation (FBC) module.
For the first module, it firstly learn representations for interests under each behavior with the paradigm of graph neural networks (capture the knowledge-aware relations between items), 
and then use the representations of knowledge-aware relations to initiate shared and behavior-specific interests for every behavior. 
For the second module, we design a GNN-based framework (capture the high-order relations between users and items)
with dynamic routing mechanism \cite{capsnet} to assign every edge on each graph (means user-item interaction under a specific behavior) to different interests. In this module, we then generate fine-grained representations for all interests by graph propagation on separate interest-level graph. And finally, we apply the self-attention mechanism on these interest representations to model the \textsl{decoupled interest-level} correlation between different behaviors for multi-behavior prediction.

To summarize, this work makes the following contributions:

\begin{itemize}
    \item We propose a novel Coarse-to-fine Knowledge-enhanced Muti-interest Learning framework (CKML) for multi-behavior recommendation, which learns shared and behavior-specific user interests for different behaviors.
    To the best of our knowledge, this is the first attempt to introduce multi-interest learning into the multi-behavior recommendation.
    \item We propose a multi-interest learning mechanism that models interest with a coarse to fine process. It contains a Coarse-grained Interest Extracting (CIE) module and a Fine-grained Behavioral Correlation (FBC) module, which better models the complex dependencies among multiple behaviors.  
    \item We conduct extensive experiments on three public datasets with a vastly different type of behaviors. The experimental results show the performance superiority and interpretability of our proposed framework.
\end{itemize}
\section{Related Work}
\label{related work}


\subsection{Multi-behavior Recommendation} 
The existing multi-behavior recommendation methods can be classified into two categories \cite{huang2021recent}. 
One is multi-behavioral representation modeling based on advanced neural networks, such as transformer and graph neural networks. 
For example, DIPN \cite{dipn} proposes a hierarchical attention network, which uses both the intra-view and inter-view attention to learn the relationships between different behaviors.
MATN \cite{matn} then uses transformer to encode the interactions of multiple behaviors.
Further, the GNN-based methods like MGNN \cite{mgnn}, MBGCN \cite{mbgcn}, GHCF \cite{ghcf} and  MBGMN \cite{mbgmn}  propose to leverage message passing on graphs to model high-order multi-behavioral interactive information.
Moreover, KHGT \cite{khgt} combines GNN and transformer together to model the global behavioral information, which not only captures the higher-order behavior between nodes, but also addresses the dynamics of behavior. 
The other category is to model different behaviors with MTL. 
DIPN \cite{dipn}, MGNN \cite{mgnn} and GHCF \cite{ghcf} regard the aggregated representations of different behaviors as shared input and simply use aggregated representations to predict each behavior individually.
NMTR \cite{nmtr}, EHCF \cite{chen2020efficient} and MBGMN \cite{mbgmn} use a transfer learning paradigm to fully interact and aggregate different behavioral representations and then make predictions separately. 
All in all, these multi-behavior methods try to capture the correlation between different behaviors, but they do not take into account the potential fine-grained interests behind each behavioral interaction. 

\subsection{Multi-interest Recommendation}
Existing methods for multi-interest learning can be divided into two paradigms. One paradigm is utilizing behavioral collaborative signals to learn multi-interest representations. For example, MIND \cite{mind} applies a dynamic routing mechanism to assign each interaction to interests. 
On this basis, Comirec \cite{comirec} leverages self-attentive mechanism to extract user interests. 
To better learn interest representations, SINE \cite{sine} and Octopus \cite{octopus} propose to model interests explicitly. 
They first build interest pools and then use attention mechanisms to explicitly activate some of the interests of users in the pool through historical user interactions. 
DGCF \cite{dgcf} introduces the dynamic routing mechanism into graphs with independence modeling among interests for multi-interest learning.
The other paradigm is leveraging structured relational information to construct multi-interest representations.
For instance, KGIN \cite{kgin} exploits knowledge graph's structural information to learn the representations of different interests.
KTUP \cite{ktup} rises a new translation-based model, which leverages implicit interests to capture the relationship between users and items.
We can find that the former does not consider the importance of knowledge-aware information in the \emph{initialization of interest clustering centers}, and the latter does not consider the importance of collaborative signals in the \emph{process of interest clustering}. 
Our method not only well \emph{initializes the interest clustering centers}, but also sufficiently utilizes the collaborative signals to assist the \emph{process of interest clustering}.
\section{Problem Definition}

\textbf{Multi-Behavior Interaction Graph.}
Let $\mathcal{U}=\{u_1, u_2, ... , u_M\}$ represent the set of users and $\mathcal{I}=\{i_1, i_2, ... , i_N\}$ represent the set of items, where $M$ and $N$ are the numbers of users and items, respectively.
In real-world recommendation scenarios, users can interact with items in multiple behaviors. 
Suppose there are $K$ types of behaviors, we denote the user-item interaction data of different behaviors as $\mathcal{Y}_{u-i} = \left\{\mathbf{Y}_{u-i}^1,\mathbf{Y}_{u-i}^2,...,\mathbf{Y}_{u-i}^K\right\}$,
where $\mathbf{Y}_{u-i}^k$ represents the interaction matrix of behavior $k$, $y_{ui}^k = 1$ denotes that user $u$ interacts with item $i$ under behavior $k$, otherwise $y_{ui}^k = 0$.
The user-item interaction data can also be regarded as a user-item bipartite graph $\mathcal{G}_{u-i}=(\mathcal{V}, \mathcal{E}_{u-i}, \mathcal{A}_{u-i}, \mathcal{T}, \mathcal{R}_{u-i})$, where $\mathcal{V} = \mathcal{U}\cup\mathcal{I}$ is the node set containing all users and items, $\mathcal{E}_{u-i} = \cup_{k \in \mathcal{R}_{u-i}}\mathcal{E}_{u-i}^{k}$ is the edge set including all behavior records between users and items.
Here $k$ denotes a specific type of behavior and $\mathcal{R}_{u-i}$ is the set of all possible behavior types.
$\mathcal{A}_{u-i} = \cup_{k \in \mathcal{R}_{u-i}}\mathbf{A}_{u-i}^{k}$ is the adjacency matrix set with $\mathbf{A}_{u-i}^{k}$ denoting adjacency matrix of a specific behavior graph $\mathcal{G}_{u-i}^{k}=(\mathcal{V},\mathcal{E}_{u-i}^{k}, \mathbf{A}_{u-i}^{k})$.

\textbf{Knowledge-aware Relation Graph.}
To explore the rich semantics of items, we define graph $\mathcal{G}_{i-i}=(\mathcal{I}, \mathcal{E}_{i-i}, \mathcal{A}_{i-i}, \mathcal{R}_{i-i})$ to leverage side information like attributes and external knowledge to depict the multi-faced characteristics of items, the definition of $\mathcal{E}_{i-i}$ and $\mathcal{R}_{i-i}$ are similar to the definition of $\mathcal{E}_{u-i}$ and $\mathcal{R}_{u-i}$, respectively.
We denote the item-item relation matrix as $\mathcal{A}_{i-i} = \left\{\mathbf{A}_{i-i}^1,\mathbf{A}_{i-i}^2,...,\mathbf{A}_{i-i}^{|\mathcal{R}_{i-i}|}\right\}$, which can be constructed with different reasons, such as items with same category, from the same restaurant, or interacted by similar users.

\textbf{Task Description.}
Generally, there is a target behavior to be optimized (e.g., purchase), which we denote as $\mathbf{Y}_{u-i}^K$, and other behaviors $\left\{\mathbf{Y}_{u-i}^1,\mathbf{Y}_{u-i}^2,...,\mathbf{Y}_{u-i}^{K-1}\right\}$ (e.g., view and tag as favorite) are treated as auxiliary behaviors for assisting the prediction of target behavior.
The goal is to predict the probability that user $u$ will interact with item $i$ under target behavior $K$.
\section{METHODOLOGY}
\begin{figure*}[t]
	\centering
	\setlength{\belowcaptionskip}{-0.2cm}
	\setlength{\abovecaptionskip}{-0.0cm}
	\includegraphics[width=0.95\textwidth]{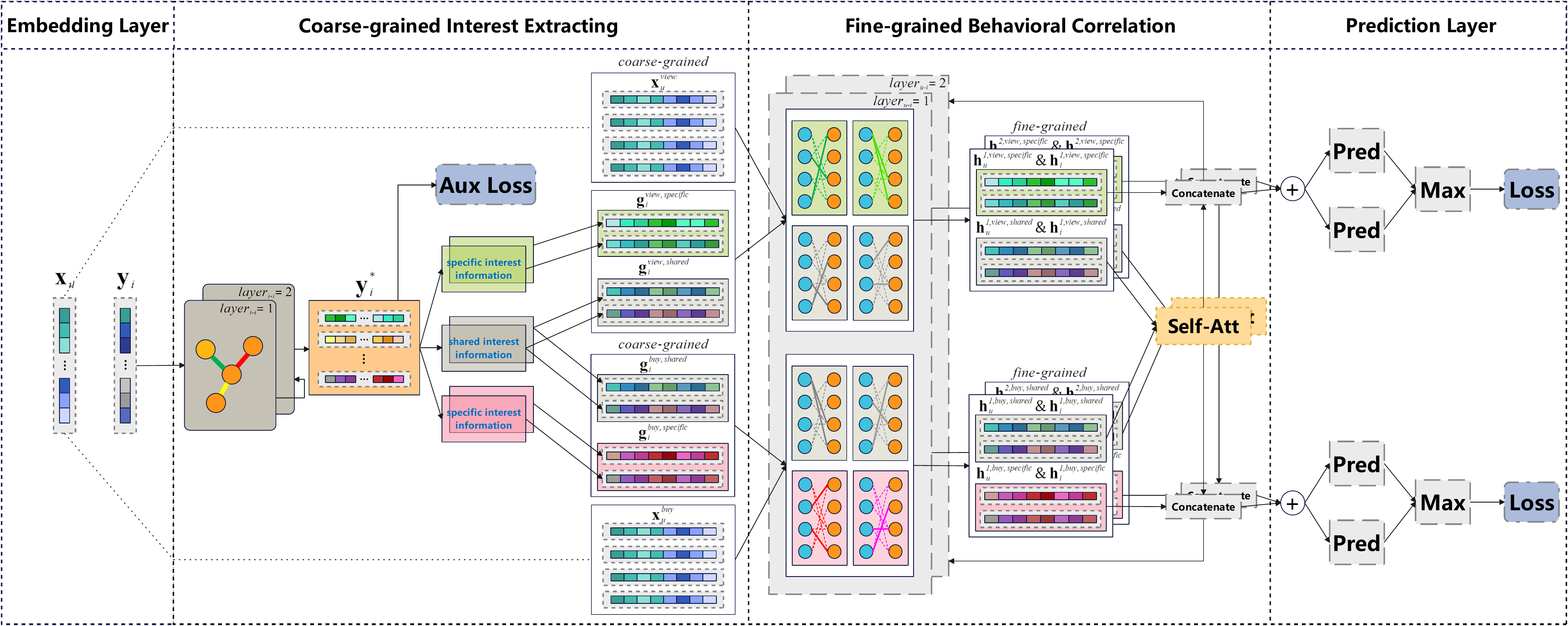}
	\caption{Illustration of the proposed CKML. For brevity, only two behaviors are represented here. The green and red rectangles represent the behavior-specific interests, while the grey rectangle represents the shared interests. The orange circles represent items, while the blue circles represent users. ($\oplus$) denotes the element-wise addition operation.}
	\label{fig:framework}
	\vspace{-3mm}
\end{figure*}
We now present the model details of our proposed CKML, which is illustrated in Figure \ref{fig:framework}. 
It consists of two core modules: (1) Coarse-grained Interest Extracting (CIE) module, which utilizes knowledge-aware relations to extract shared and behavior-specific interests for multiple behaviors; and (2) Fine-grained Behavioral Correlation (FBC) module, which allocates different interactions to different interests under each behavior, then models the complex behavior dependencies with interest-aware correlations. 


\subsection{Embedding Layer}
In industrial applications, users and items are often denoted as high-dimensional one-hot vectors.
Generally, given a user-item pair $(u, i)$, we apply the embedding lookup operation for user $u$ and item $i$ to obtain the embedding vectors:
\begin{equation}
\setlength{\abovedisplayskip}{0.5pt}
\setlength{\belowdisplayskip}{0.5pt}
\mathbf{x}_u = \mathbf{E}_u^{T} \cdot \mathbf{p}_u, \ 
\mathbf{y}_i = \mathbf{E}_i^{T} \cdot \mathbf{p}_i
\end{equation}
where $\mathbf{E}_u \in \mathbb{R}^{M \times d}$ and $\mathbf{E}_i \in \mathbb{R}^{N \times d}$ are the created embedding tables for users and items, $\mathbf{p}_u \in \mathbb{R}^{M}$ and $\mathbf{p}_i \in \mathbb{R}^{N}$ denotes the one-hot IDs of user $u$ and item $i$, and $d$ is the embedding size. 


\subsection{Coarse-grained Interest Extracting}
Knowledge-aware item-item relations are widely used to supplement semantic information and assist representation learning \cite{khgt,kgin,ktup}. 
Inspired by the strong semantics of relations in the knowledge-aware relation graph \cite{kgin,kgat,khgt}, we propose a Coarse-grained Interest Extraction (CIE) module to extract users' interests which motivates users' interactions of multiple behaviors. 
In this way, we obtain the initial interest clustering centers.
There are two main components in CIE: the first part is the knowledge-aware relation modeling which is designed for capturing the semantic information from the knowledge-aware item-item relation graph, while the second part is the behavior-aware interest extraction which is designed to utilize the semantic information obtained in the previous part to make an extraction of interests. 

\subsubsection{Knowledge-aware Relation Modeling}
\label{sec:KIRM}
Most existing multi-interest methods initialize interests with random generated vectors \cite{mind,dgcf}, which fails to endow interests with semantics and may lead to a chaotic interest division. 
Since we have emphasized the importance of initializing interest clustering centers in Section \ref{intro}, and inspired by the knowledge graph based methods \cite{kgin,ktup,khgt}, we use knowledge-aware information to initialize interest representations.
Thanks to its high capability in modeling relational data and great performance in representation learning, we seek to utilize knowledge-aware relations for interest extraction under the paradigm of graph neural networks in this component.
Specially, we firstly partition the knowledge-aware relation graph $\mathcal{G}_{i-i}$ into several relation-specific sub-graphs 
$\mathcal{G}_{i-i}^{1}, \mathcal{G}_{i-i}^{2},...,\mathcal{G}_{i-i}^{|\mathcal{R}_{i-i}|}$, and the
corresponding adjacency matrices are $\mathbf{A}_{i-i}^{1},\mathbf{A}_{i-i}^{2},...,\mathbf{A}_{i-i}^{|\mathcal{R}_{i-i}|}$.
As for the message propagation, we adopt the state-of-the-art GCN models, such as LightGCN \cite{lightgcn}, LR-GCCF \cite{lr-gccf}, GCN \cite{gcn} and NGCF \cite{ngcf}, for graph information aggregation. And the neighbor propagation process in each layer of each sub-graph can be formulated as:
\begin{equation}
\setlength{\abovedisplayskip}{0.5pt}
\setlength{\belowdisplayskip}{0.5pt}
\mathbf{y}_{i}^{r, l}=\mathop{Agg}\limits_{j \in N_{i}}\left(\mathbf{y}_{j}^{r, l-1}, \mathbf{A}_{i-i}^{r}\right)
\end{equation}
where $r$ denotes the type of relation, $l$ denotes the layer of GNN, $N_{i}$ denotes the neighbors of item $i$, and $\mathbf{y}_{i}^{r, 0} = \mathbf{y}_{i}$ is the initial embedding for item $i$. 
After the propagation, we average the generated representations from all layers to get the final representations: 
\begin{equation}
\setlength{\abovedisplayskip}{0.5pt}
\setlength{\belowdisplayskip}{0.5pt}
\mathbf{y}_{i}^{r} = {\sum\limits_{l=0}^{L_{i-i}}{\mathbf{y}_{i}^{r, l}}}/{(L_{i-i}+1)}
\end{equation}
where $\mathbf{y}_{i}^{r} \in \mathbb{R}^{1 \times d}$ and $L_{i-i}$ is the number of GNN layers setting for modeling the knowledge-aware relation graph.
We use the same number of layers for all relations here for simplicity.
\subsubsection{Behavior-aware Interest Extraction.}
\label{sec:BIE}
Since we have obtained representations for all relations, how to effectively extract interests for different behaviors remains a challenge.
As shown in Figure \ref{fig:relation}, different behaviors exhibit diverse interest patterns.
Some interests are shared across multiple behaviors, while others are unique for specific behaviors.
This is similar to shared expert information and specific expert information under different tasks in multi-task learning.
Motivated by the customized gate presented in PLE \cite{tang2020progressive} which achieves great performance in multi-task learning,  
we creatively propose to introduce shared interest and behavior-specific interest for multi-interest learning. 
The shared interest is designed to correlate with other types of behaviors at the level of interest, which can better leverage the potential complementary information of same interest within multiple behaviors. 
And the specific interest decouples and retains the independence of the corresponding behaviors, thus alleviating the influence of noise.
We first combine the representations of all relations into a unified vector:
\begin{equation}
\setlength{\abovedisplayskip}{0.5pt}
\setlength{\belowdisplayskip}{0.5pt}
\mathbf{y}_{i}^{*} = \mathop{Concatenate}\limits_{r \in \mathcal{R}_{i-i} }{\mathbf{y}_{i}^{r}},
\end{equation}

After that, we use a non-linear transformation which is generally used to model the combinations among relations to convert relations into multiple interests. 
For the specific interests, we have:
\begin{equation}
\setlength{\abovedisplayskip}{0.5pt}
\setlength{\belowdisplayskip}{0.5pt}
\mathbf{g}_{i}^{k, spe}=\mathop{Concatenate}\limits_{s=1}^{N_{spe}}\left(\mathop{LeakyReLU}\left(\mathbf{y}_{i}^{*}\cdot\mathbf{W}_{s}^{k}+\mathbf{b}_{s}^{k}\right)\right)
\end{equation}
where $N_{spe}$ is the number of specific interests for each behavior, $s$ is the $s$-$th$ interest, $\mathbf{W}_{s}^{k} \in \mathbb{R}^{(|\mathcal{R}_{i-i}| * d) \times ({d\over{N_{spe}}})}$ and $\mathbf{b}_{s}^{k} \in \mathbb{R}^{1 \times ({d\over{N_{spe}}})}$ are transformation matrix and bias matrix, and $\mathbf{g}_{i}^{k, spe}$ denotes the extracted behavioral-specific interests for behavior $k$. 
Notice that we use $1\over{N_{spe}}$ of the original item embedding size as the interest size to keep similar space usage as single-interest models, and we apply the same compressed form to shared interests.
For the behavioral shared interests, we have:
\begin{equation}
\setlength{\abovedisplayskip}{0.5pt}
\setlength{\belowdisplayskip}{0.5pt}
\mathbf{g}_{i}^{k, sha}=\mathop{Concatenate}\limits_{s=1}^{N_{sha}}\left(\mathop{LeakyReLU}\left(\mathbf{y}_{i}^{*}\cdot\mathbf{W}_{s}+\mathbf{b}_{s}\right)\right)
\end{equation}
where $N_{sha}$ is the number of shared interests, $s$ is the $s$-$th$ interest, $\mathbf{W}_{s} \in \mathbb{R}^{(|\mathcal{R}_{i-i}| * d) \times ({d\over{N_{sha}}})}$ and $\mathbf{b}_{s} \in \mathbb{R}^{1 \times ({d\over{N_{sha}}})}$ are transformation matrix and bias matrix, $\mathbf{g}_{i}^{k, sha}$ denotes the extracted interests for behavior $k$. Since the parameters of different behaviors are shared, the shared representations of different $k$ in this equation are consistent.

Finally, we union the representations of shared and specific interests as the output of CIE:
\begin{equation}
\setlength{\abovedisplayskip}{0.5pt}
\setlength{\belowdisplayskip}{0.5pt}
\mathbf{g}_{i}^{k}=\mathbf{g}_{i}^{k, spe}\|\mathbf{g}_{i}^{k, sha}
\end{equation}
where $(\|)$ is the concatenation operation between two vectors. 
For convenience, we set ${d\over{N_{spe}}} = {d\over{N_{sha}}} = d^{*}$, $N_{*} = N_{spe}+N_{sha}$. 
\begin{algorithm}[t]
    \SetKwInOut{Input}{Input}
    \SetKwInOut{Output}{Output}
    \Input{$\mathbf{A}_{u-i}^{k} \in \mathbb{R}^{(M+N)\times(M+N)}$,
    $\forall u \in \mathcal{U}, \forall i \in \mathcal{I}:$ \\ $ \mathbf{x}_{u}^{l,k},\mathbf{g}_{i}^{l,k},\mathbf{t}_{u}^{k},\mathbf{t}_{i}^{k} \in \mathbb{R}^{N_{*}\times d^{*}}$}
	\Output{Vector representations~$\mathbf{h}_u^{l,k}$ for all~$u \in \mathcal{U}$ and~$\mathbf{h}_i^{l,k}$ for all~$i \in \mathcal{I}$}
	\SetAlgorithmName{Algorithm}{} \\
{\color{blue}\tcc{Step 1. Initialize.}}
	   $\mathcal{E}_{u-i}^{k} \leftarrow \mathbf{A}_{u-i}^{k}$,
	   $\mathbf{a}_{0}^{l,k} \leftarrow \mathbf{1}^{N_{*} \times |\mathcal{E}_{u-i}^{k}|}$
	   \\
	   $\mathbf{h}_{u,0}^{l,k} \leftarrow \mathbf{x}_{u}^{l,k} + \mathbf{t}_{u}^{k}$,
	   $\mathbf{h}_{i,0}^{l,k} \leftarrow \mathbf{g}_{i}^{l,k} + \mathbf{t}_{i}^{k}$
	   \\
     \For{$t=1$ to $N_{iter}$}{
{\color{blue}\tcc{Step 2. Normalize coefficients.}}
        $\mathbf{a}_{t}^{l,k} \leftarrow \mathop{Softmax}(\mathbf{a}_{t-1}^{l,k}/\tau)$\\
          \For{$s=1$ to $N_{*}$}
           {
{\color{blue}\tcc{Step 3. Embedding propagation.}}
            $\mathbf{h}_{u,t}^{l,k}[s] \leftarrow \sum\limits_{(u,i) \in \mathcal{E}_{u-i}^{k}}  {\mathbf{a}_{t}^{l,k}[s,(u,i)] \cdot \mathbf{h}_{i,0}^{l,k}[s]\over{\sum\limits_{(u,i) \in \mathcal{E}_{u-i}^{k}}\mathbf{a}_{t}^{l,k}[s,(u,i)]}}, \forall u \in \mathcal{U}$ \\
            $\mathbf{h}_{i,t}^{l,k}[s] \leftarrow \sum\limits_{(i,u) \in \mathcal{E}_{u-i}^{k}} {\mathbf{a}_{t}^{l,k}[s,(i,u)] \cdot \mathbf{h}_{u,0}^{l,k}[s]\over{\sum\limits_{(i,u) \in \mathcal{E}_{u-i}^{k}}\mathbf{a}_{t}^{l,k}[s,(i,u)]}}, \forall i \in \mathcal{I}$ 
{\color{blue}\tcc{Step 4. Update coefficients.}}
           $\mathbf{a}_{t}^{l,k}[s] \leftarrow \mathbf{a}_{t-1}^{l,k}[s]+$\\ $\mathop{Stack}\limits_{(p,q) \in \mathcal{E}_{u-i}^{k}}({\mathbf{h}_{q,0}^{l,k}[s]\over{|\mathbf{h}_{q,0}^{l,k}[s]|}} \cdot \mathop{tanh}({\mathbf{h}_{p,t}^{l,k}[s] \over {|\mathbf{h}_{p,t}^{l,k}[s]|}})^{\top})$ \\
           }
     }   
{\color{blue}\tcc{Step 5. Information aggregation.}}
$\mathbf{h}_u^{l,k} \leftarrow \mathop{Agg}(\mathop{Concatenate}\limits_{s=1}^{N_{*}}(\mathbf{h}_{u,N_{iter}}^{l,k}[s]), \mathbf{A}_{u-i}^{k}), \forall u \in \mathcal{U}$ \\  
$\mathbf{h}_i^{l,k} \leftarrow \mathop{Agg}(\mathop{Concatenate}\limits_{s=1}^{N_{*}}(\mathbf{h}_{i,N_{iter}}^{l,k}[s]), \mathbf{A}_{u-i}^{k}), \forall i \in \mathcal{I}$\\
\caption{The allocation of interests for the $k$-$th$ behavior on the $l$-$th$ layer.}

\label{alg:1}
\end{algorithm}
\vspace{-4mm}

\subsection{Fine-grained Behavioral Correlation}
Existing multi-behavior methods model the dependencies among multiple behaviors without distinguishing the diverse interests on which different interactions are based, which may inevitably introduce noise if the interactions are due to different interests.

In the previous part, we have preliminarily extracted the potential interest of items based on the knowledge-aware relations. 
However, this is only a node-wise partitioning, and does not divide specific interactions (i.e. edges on the graph) into interests. 
Here "node-wise" means the level of users and items, while the corresponding "edge-wise" denotes a finer-grained level that considers each interaction between users and items.
To address this problem, we propose a Fine-grained Behavioral Correlation (FBC) layer to further allocate each interaction to different interests and model the dependence between behaviors at the level of interest.
FBC is composed of two key components: The first one is interest-aware behavior allocation which is designed to further allocate each interaction to different interests. And the second one is interest-aware dependence modeling which is designed to capture inter-behavioral correlations and adequately leverage this information at each layer.

\subsubsection{Interest-aware Behavior Allocation.}

To allocate the edges on the graph $\mathcal{G}_{u-i}$ to different interests under each behavior, we apply the disentangled representation learning \cite{dgcf,mind,comirec} for behavior allocation.
We firstly partition the provided multi-behavior user-item graph $\mathcal{G}_{u-i}$ into behavior-specific sub-graphs $\mathcal{G}_{u-i}^{1}, \mathcal{G}_{u-i}^{2},...,\mathcal{G}_{u-i}^{K}$, and the
corresponding adjacency matrices are $\mathbf{A}_{u-i}^{1},\mathbf{A}_{u-i}^{2},...,\mathbf{A}_{u-i}^{K}$, which can be formulated as:
\begin{equation}
\setlength{\abovedisplayskip}{0.5pt}
\setlength{\belowdisplayskip}{0.5pt}
\mathbf{A}_{u-i}^{k}=\left(\begin{array}{cc}
0 & \mathbf{Y}_{u-i}^{k} \\
\left(\mathbf{Y}_{u-i}^{k}\right)^{T} & 0
\end{array}\right)
\end{equation}
where $\mathbf{Y}_{u-i}^{k}$ is the user-item adjacency interaction matrix of behavior $k$, $\mathbf{A}_{u-i}^{k} \in \mathbb{R}^{(M+N)\times(M+N)}$, $M$ and $N$ denote the number of users and items, respectively.
As for the processing of time, we simply follow KHGT \cite{khgt}, and initialize the embedding of time as  $\mathbf{t}_{u-i}^{k} \in \mathbb{R}^{N_{*}\times d^{*}}$ for each interaction.


To better illustrate the process of the allocation of interests on each layer, we take the $k$-$th$ behavior on the $l$-$th$ layer as an example.
As shown in Algorithm \ref{alg:1}, we use $\mathcal{E}_{u-i}^{k} = \left\{(p,q)|\mathbf{A}_{u-i}^{k}[p,q] \neq 0 \right\} $ to represent the set of edges on graph $\mathcal{G}_{u-i}^{k}$.
Meanwhile, we set $\mathbf{a}_{0}$ as the initial weight for each edge on $\mathcal{G}_{u-i}^{k}$ and initialize the embedding for each user and item \textcolor{blue}{(Step 1)}. And for detail, we have $\mathbf{x}_{u}^{0, k} = \mathbf{x}_{u}^{k}$ and $\mathbf{g}_{i}^{0, k} = \mathbf{g}_{i}^{k}$.
Next, we start iterative process. 
In the $t$-$th$ iteration, in order to get distributions across all interests, we use the softmax function to normalize these coefficients \textcolor{blue}{(Step 2)}:
\begin{equation}
\setlength{\abovedisplayskip}{0.5pt}
\setlength{\belowdisplayskip}{0.5pt}
\mathbf{a}_{t}^{l,k}(s,(u,i))=\frac{\exp {(\mathbf{a}_{t-1}^{l,k}(s,(u,i)))/{\tau})}}{\sum_{s=1}^{N_{*}} \exp {(\mathbf{a}_{t-1}^{l,k}(s,(u,i))/{\tau})}}
\end{equation}
where $\mathbf{a}_{t}^{l,k}$ denotes the vector of weight coefficients of each edge on the $l$-$th$ layer of graph $\mathcal{G}_{u-i}^{k}$ in the $t$-$th$ iteration, $\tau$ is the temperature coefficient, $\forall (u,i) \in \mathcal{E}_{u-i}^{k}$, $s$ denotes the $s$-$th$ interest.
Furthermore, in each iteration, we assign all the edges on the graph $\mathcal{G}_{u-i}^{k}$ to each interest of users and items on the graph \textcolor{blue}{(Step 3)}.
At this step, $\mathbf{h}_{u,t}^{l,k}[s]$ and $\mathbf{h}_{i,t}^{l,k}[s]$ represent the $s$-$th$ interest for user $u$ and item $i$ after the allocation of the edge weights, respectively.
Last but not least, we calculate the affinity between each pairs of nodes on the graph $\mathcal{G}_{u-i}^{k}$ to update the weight of each edge \textcolor{blue}{(Step 4)}. Here, $\mathbf{a}_{t}[s]$ denotes the updated weight of edges at the $t$-$th$ iteration of the $s$-$th$ interest.
After all of the iterations, we finally take the representation generated by the last iteration as the final output, and aggregate them with GCN models \textcolor{blue}{(Step 5)}, which is the same as the aggregators in Section \ref{sec:KIRM}.


\subsubsection{Interest-aware Behavioral Correlation.}
After the allocation of interests for each node on each layer, we need to correlate information between behaviors at the interest level.
And we just correlate the information between the representations of shared interests of each behavior with a self-attention network \cite{vaswani2017attention} because the behavior-specific interests contain few useful information for the target behavior and may contain noise. 
For instance, in the Yelp and MovieLens-10M datasets, there are behaviors (Dislike) that are contrary to the target behavior (Like), which may interfere with the learning of target behavior. 
For better convergence, we apply residual connection to the output of self-attention \cite{resnet}, which can be formulated as:
\begin{equation}
\setlength{\abovedisplayskip}{0.5pt}
\setlength{\belowdisplayskip}{0.5pt}
\left\{\begin{array}{c}
\begin{aligned}
\tilde{\mathbf{h}}_{n}^{l,k,sha} &=\mathbf{M H}-\operatorname{Att}\left(\mathbf{h}_{n}^{l,k,sha}\right)+\sum\limits_{k^{\prime}=1}^{K}\mathbf{h}_{n}^{l,k^{\prime},sha}
\\\mathbf{M H}-\operatorname{Att}\left(\mathbf{h}_{n}^{l,k,sha}\right)&=\mathop{Concatenate}\limits_{h=1}^{H} (\sum_{k^{\prime}=1}^{K} \lambda_{k, k^{\prime}}^{n, h} \cdot \tilde{\mathbf{V}}^{h} \cdot \mathbf{h}_{n}^{l,k^{\prime},sha}) \\
\lambda_{k, k^{\prime}}^{n, h} &=\frac{\exp \bar{\lambda}_{k, k^{\prime}}^{n, h}}{\sum_{k^{\prime}=1}^{K} \exp \bar{\lambda}_{k, k^{\prime}}^{n, h}} \\ \bar{\lambda}_{k, k^{\prime}}^{n, h}&=\frac{\left(\tilde{\mathbf{Q}}^{h} \cdot \mathbf{h}_{n}^{l,k,sha}\right)^{\top}\left(\tilde{\mathbf{K}}^{h} \cdot \mathbf{h}_{n}^{l,k^{\prime},sha}\right)}{\sqrt{d^{*} / H}}
\end{aligned}
\end{array}\right.
\end{equation}
where $\tilde{\mathbf{Q}}^{h}$, $\tilde{\mathbf{K}}^{h}$, $\tilde{\mathbf{V}}^{h}$  $\in \mathbb{R}^{{d^{*}\over{H}}\times {d^{*}\over{H}}}$ are learnable projection matrices of the $h$-$th$ head. $n \in \mathcal{U}\cup\mathcal{I}$ denotes the user $u$ or item $i$, $\lambda_{k, k^{\prime}}^{n, h}$ represents the relevance score between $k$-$th$ and $k^{\prime}$-$th$ behaviors of the $h$-$th$ head for the node $n$. For the information propagation of the $k$-$th$ behavior, we have:
\begin{equation}
\setlength{\abovedisplayskip}{0.5pt}
\setlength{\belowdisplayskip}{0.5pt}
\left\{\begin{array}{c}
\begin{aligned}
    \mathbf{x}_{u}^{l+1,k} &= \mathbf{h}_{u}^{l,k,spe} \| \tilde{\mathbf{h}}_{u}^{l,k,sha}+\mathbf{x}_{u}^{l,k}, \forall u \in \mathcal{U}\\
    \mathbf{g}_{i}^{l+1,k} &= \mathbf{h}_{i}^{l,k,spe} \| \tilde{\mathbf{h}}_{i}^{l,k,sha}+\mathbf{g}_{i}^{l,k}, \forall i \in \mathcal{I} 
\end{aligned}
\end{array}\right.
\end{equation}
where $l \in [0,1,...,L_{u-i}]$, $L_{u-i}$ is the number of GNN layer, $(\|)$ is the concatenated operation for two vectors.


\subsection{Joint Optimization}
\subsubsection{The Prediction of the U-I Interaction}
\label{sec:u-i pred}
In the above parts, we have obtained the output $\mathbf{h}_{u}^{l,k,spe}$ and $\tilde{\mathbf{h}}_{u}^{l,k,sha}$, $\forall l \in [1,2,...,L_{u-i}], \forall k \in [1,2,...,K], \forall u \in \mathcal{U}$, similar operations are applied for the item $i$. To aggregate the information of each layer, we follow KHGT \cite{khgt}, and simply add them up. The unified formulation is:
\begin{equation}
\setlength{\abovedisplayskip}{0.5pt}
\setlength{\belowdisplayskip}{0.5pt}
\begin{aligned}
    \mathbf{h}_{v}^{*,k} &= \sum\limits_{l=1}^{L_{u-i}} (\mathbf{h}_{v}^{l,k,spe} \| \tilde{\mathbf{h}}_{v}^{l,k,sha}), \forall v \in \mathcal{U} \cup \mathcal{I} \\
\end{aligned}
\end{equation}
where $\mathbf{h}_{v}^{*,k} \in \mathbb{R}^{N_{*}\times d^{*}}$, $k$ represents the $k$-$th$ behavior.
Inspired by ComiRec \cite{comirec}, we make separate predictions for each interest under each behavior and take the maximum of all the predictions under each behavior, which can be formulated as:
\begin{equation}
\setlength{\abovedisplayskip}{0.5pt}
\setlength{\belowdisplayskip}{0.5pt}
    \hat{\mathbf{o}}_{u,i}^{k} = \max\limits_{s=1}^{N_{*}}(\sum\limits_{j}^{d^{*}}(\mathbf{h}_{u}^{*,k}[s] \circ \mathbf{h}_{i}^{*,k}[s])[j])\\
\end{equation}
where $s \in [1,2,...,N_{*}]$ denotes the $s$-$th$ interest, ($\circ$) is the hadamard product operation.

Finally, to perform the model optimization, we follow KHGT \cite{khgt} and use marginal pair-wise \textit{Bayesian Personalized Ranking} (BPR) loss function to minimize the following loss function:
\begin{equation}
\setlength{\abovedisplayskip}{0.5pt}
\setlength{\belowdisplayskip}{0.5pt}
\mathcal{L}_{u-i}=\sum_{k=1}^{K} \sum_{(u,p,q)\in \mathcal{O}_{u-i,k}} \alpha^{k}*\max \left(0,1-\hat{\mathbf{o}}_{u,p}^{k}+\hat{\mathbf{o}}_{u,q}^{k}\right)
\end{equation}
where the $\alpha^{k} \in [0,1]$ denotes the coefficient of loss for each behavior, $\mathcal{O}_{u-i,k} = \left\{(u,p,q)|(u,p)\in \mathcal{O}_{u-i,k}^{+}, (u,q) \in \mathcal{O}_{u-i,k}^{-} \right\}$ denotes the training dataset. $\mathcal{O}_{u-i,k}^+$ indicates observed positive user-item interactions under behavior $k$ and $\mathcal{O}_{u-i,k}^-$ indicates unobserved user-item interactions under behavior $k$. 

\subsubsection{The Prediction of the Knowledge-aware Item-item Relation}

Inspired by self-supervised learning on graphs \cite{ssl-graph}, we use the information of item-item relations to reconstruct the item-item graphs, which can be considered as a self-supervised relation reconstruction task to enhance the learning of interest representations. 

In detail, since we obtained the representations of each relation for all item $i \in \mathcal{I}$, i.e. $\mathbf{y}_{i}^{*}$ at Section \ref{sec:KIRM}, we calculate prediction scores for each relation between items: 
\begin{equation}
\setlength{\abovedisplayskip}{0.5pt}
\setlength{\belowdisplayskip}{0.5pt}
    \hat{\mathbf{o}}_{i,i^{\prime}}^{r} = \sum\limits_{j}^{d}(\mathbf{y}_{i}^{r} \circ \mathbf{y}_{i^{\prime}}^{r})[j]\\
\end{equation}
where $r$ represents the $r$-$th$ relation.
We then use the BPR loss to reconstruct the graph $\mathcal{G}_{i-i}^{r}$, which can be formulated as:
\begin{equation}
\setlength{\abovedisplayskip}{0.5pt}
\setlength{\belowdisplayskip}{0.5pt}
\mathcal{L}_{i-i}=-\sum_{r=1}^{|\mathcal{R}_{i-i}|} \sum_{(i, p, q) \in O_{i-i,r}} \ln \sigma\left(\hat{o}_{i, p}^{r}-\hat{o}_{i, q}^{r}\right)
\end{equation}
where $\mathcal{O}_{i-i,r} = \left\{(i,p,q)|(i,p)\in \mathcal{O}_{i-i,r}^{+}, (i,q) \in \mathcal{O}_{i-i,r}^{-} \right\}$ denotes the training dataset of the item-item relation graph reconstructive task, which is similar to the definition in Section \ref{sec:u-i pred}. 
Finally, for the total loss, we have:
\begin{equation}
\setlength{\abovedisplayskip}{0.5pt}
\setlength{\belowdisplayskip}{0.5pt}
    \mathcal{L}_{total} = \mathcal{L}_{u-i}+\beta\mathcal{L}_{i-i}+\lambda\|\Theta\|_{\mathrm{F}}^{2}
\end{equation}
where $\Theta$ represents the set of all trainable parameters, $\lambda$ is the weight for the regularization term, $\beta \in [0,1]$ is the weight of $\mathcal{L}_{i-i}$. 


\subsection{Complexity Analysis}

\subsubsection{Time Complexity.}In CIE, we spend $\mathcal{O}( L_{i-i}\left|\mathcal{E}_{i-i}\right| d)$ for message propagation in the knowledge-aware item-item graph, where $L_{i-i}$ denotes the number of GNN layers in handling item-item relations, $\left|\mathcal{E}_{i-i}\right|$ is the number of edges on $\mathcal{G}_{i-i}$ and $d$ is the embedding size. After that, the time spent to extract the interest from the item-item relation is $\mathcal{O}(\left|\mathcal{R}_{i-i}\right| d^{2})$, where $\left|\mathcal{R}_{i-i}\right|$ refers to the number of relations. In FBC, it takes $\mathcal{O}(L_{u-i}\left|\mathcal{E}_{u-i}\right| d)$ to propagate embedding in the user-item bipartite graph, where $L_{u-i}$ is the number of GNN layers in handling user-item relations and $\left|\mathcal{E}_{u-i}\right|$ denotes the number of edges on $\mathcal{G}_{u-i}$. Besides, the computational complexity of the self-attention mechanism is $\mathcal{O}(K L_{u-i} d^{2})$, where $K$ is the number of behaviors. In summary, the overall time complexity of CKML mainly comes from the GNN part. The time complexity of our model is comparable to other GNN-based methods and we perform experiment to validate it in Section \ref{performance}.

\subsubsection{Space Complexity.} Most of the parameters that the model needs to learn are the embedding of the user and item, which costs $\mathcal{O}((M+N)*d)$. The space size of the transformation matrixs in extracting shared interests and specific interests are $\mathcal{O}(\left|\mathcal{R}_{i-i}\right| d^{2}+d)$ and $\mathcal{O}(K\left|\mathcal{R}_{i-i}\right| d^{2}+Kd)$, where $K$ is the number of behaviors. The space size of $\tilde{\mathbf{Q}}^{h}$, $\tilde{\mathbf{K}}^{h}$ and $\tilde{\mathbf{V}}^{h}$ in the attentional mechanism requires $\mathcal{O}(L_{u-i}*{(d^{*})^{2}\over{H}})$, where the $d^{*} = {d\over{N_{spe}}} = {d\over{N_{sha}}}$, $H$ is the number of attention head. All in all, CKML have limited additional parameters except for the embedding of the user and item. 


\section{Experiments}
\label{experiments}

    
    
    
    


\subsection{Experimental Setting}
\subsubsection{Dataset Description}

Closely following KHGT \footnote{https://github.com/akaxlh/KHGT} \cite{khgt}, we evaluate our model on three datasets (i.e., \textbf{Yelp}, \textbf{MovieLens-10M}, and \textbf{Online Retail}) with the same parameter settings and preprocessing as KHGT.
The behavior types and statistics of the three datasets are shown in Table \ref{tab:dataset}.

\subsubsection{Evaluation Protocols}
We apply two widely used metrics i.e. Hit Ratio (HR@$N$) and Normalized Discounted Cumulative Gain (NDCG@$N$) to evaluate the performance and we set $N=10$ by default in all experiments. 
To fairly compare our models and baselines, we follow the evaluation settings of KHGT. 
Specifically, the last interacted item on the behavior to be predicted is used as a positive example in the test data, while the 99 randomly selected items the user has not interacted with are taken as negative examples.

\begin{table}[t]
\setlength{\abovecaptionskip}{0cm}
\setlength{\belowcaptionskip}{0cm}
\caption{Statistics of evaluation datasets.}
\centering
\resizebox{\linewidth}{!}{
\begin{tabular}{c|ccccc}
\toprule
Dataset & \#User & \#Item & \#Interaction & \#Target Interaction &  \#Interactive Behavior Type \\ \hline
Yelp & 19,800 & 22,734 & $1.4 \times 10^6$ & 677,343 & \{Tip,Dislike,Neutral,Like\} \\ \hline
MovieLens & 67,788 & 8,704 & $9.9 \times 10^6$ & 4,970,984 & \{Dislike,Neutral,Like\} \\ \hline
Online Retail & 147,894 & 99,037 & $7.7 \times 10^6$ & 642,916 & \{Page View, Favorite, Cart, Purchase\} \\  \bottomrule

\end{tabular}
}
\label{tab:dataset}
\vspace{-5mm}
\end{table}
\subsubsection{Baseline Models}
To verify the effectiveness of our CKML model, we compare it with various baseline models, which can be categorized into four groups: (A) Single-behavior non-graph models (BiasMF \cite{biasmf}, DMF \cite{dmf}, AutoRec \cite{autorec}); (B) Single-behavior graph models (ST-GCN \cite{stgcn}, NGCF \footnote{https://github.com/xiangwang1223/neural\_graph\_collaborative\_filtering} \cite{ngcf}, DGCF \cite{dgcf}, KGAT \cite{kgat}); (C) Multi-behavior non-graph models (NMTR \cite{nmtr}, DIPN \cite{dipn}, MATN\cite{matn}); (D) Multi-behavior graph models (NGCF$_{M}$ \cite{ngcf}, LightGCN$_{M}$ \footnote{https://github.com/kuandeng/LightGCN} \cite{lightgcn}, MBGCN \cite{mbgcn}, KHGT \cite{khgt}).

\subsubsection{Parameter Settings}
Our proposed CKML is implemented in TensorFlow \cite{TensorFlow}. We fix the embedding size as 16 in line with KHGT for a fair comparison. The batch size is searched in \{16,32,64\}. We initialize the parameters using Xavier \cite{xavier}. The parameters are optimized by Adam \cite{adam}, while the learning rate and decay rate are set to $10^{-3}$ and 0.96, respectively. We search the number of GNN layers in \{1,2,3,4\} for the knowledge-aware item-item graph and user-item bipartite graph, respectively. We set the number of self-attention head to 2. The number of shared interests is varied in \{1,2,4\} as well as the number of specific interests, which is investigated in Sections \ref{section:num_interest}. 
The temperature coefficient used in the interest-aware behavior allocation is tuned in \{0.1,1,5,10,20\}, and the corresponding number of iterations is set to 2. 
We conduct a grid search of the loss coefficient for each behavior in \{0,0.2,0.4,0.6,0.8,1\}. All experiments are run for 5 times and average results are reported.

\subsection{Performance Comparison}
\label{performance}
\subsubsection{Effectiveness Comparison}
Table \ref{comparisons_model} shows the performance of different methods on three datasets with respect to HR@10 and NDCG@10. 
We have the following findings:

\textbf{The effectiveness of CKML model.} 
Our proposed CKML consistently achieves the best results on all datasets.
More specifically, CKML improves the strongest baselines by \textbf{1.82\%}, \textbf{6.67\%} and \textbf{11.65\%} in terms of HR (\textbf{3.48\%}, \textbf{14.91\%} and \textbf{16.19\%} in terms of NDCG) on Yelp, MovleLens and Retail datasets, respectively.
The great improvements over baselines demonstrate the effectiveness of CKML for multi-behavior recommendation.

\begin{table}[t]
   \setlength{\abovecaptionskip}{0cm}
   \setlength{\belowcaptionskip}{-0.0cm}
\caption{The overall performance comparison. Boldface denotes the highest score and underline indicates the results of the  best baselines. $\star$ represents significance level $p$-value $<0.05$ of comparing CKML with the best baseline.}
    \centering
    \begin{threeparttable}
	\resizebox{\linewidth}{!}{
    \begin{tabular}{c|cccccccc}
    \toprule
    \multirow{2}{*}{Model}&
    \multicolumn{2}{c}{Yelp}&\multicolumn{2}{c}{MovieLens}&\multicolumn{2}{c}{Retail}\cr
    \cmidrule(lr){2-3} \cmidrule(lr){4-5} \cmidrule(lr){6-7} 
    &HR&NDCG&HR&NDCG&HR&NDCG\cr
    \midrule
    BiasMF&0.755&0.481&0.767&0.490&0.262&0.153\cr
    DMF&0.756&0.485&0.779&0.485&0.305&0.189\cr
    AutoRec&0.765&0.472&0.658&0.392&0.313&0.190\cr
    \hline
    ST-GCN&0.775&0.465&0.738&0.444&0.347&0.206\cr
    NGCF&0.789&0.500&0.790&0.508&0.302&0.185\cr
    DGCF&0.861&0.587&0.827&0.499&0.194&0.101\cr
    KGAT&0.835&0.543&0.817&0.514&0.377&0.214\cr
    \hline
    NMTR&0.790&0.478&0.808&0.531&0.332&0.179\cr
    DIPN&0.791&0.500&0.811&0.540&0.317&0.178\cr
    MATN&0.826&0.530&0.847&0.569&0.354&0.209\cr
    \hline
    NGCF$_{M}$& 0.793 & 0.492 & 0.825 & 0.546 & 0.374 & 0.221\cr
    LightGCN$_{M}$ & 0.873 & 0.573 & \underline{0.869} & 0.595 & \underline{0.472} & 0.277 \cr 
    MBGCN&0.796&0.502&0.826&0.553&0.369&0.222\cr
    KHGT &\underline{0.880}&\underline{0.603}&0.861&\underline{0.597}&0.464&\underline{0.278}\cr
    \hline
    \textbf{CKML}&\textbf{0.896}$^\star$&\textbf{0.624}$^\star$&\textbf{0.927}$^\star$&\textbf{0.686}$^\star$&\textbf{0.527}$^\star$&\textbf{0.323}$^\star$&\cr
    \hline
    Rel Impr.&1.82\%&3.48\%&6.67\%&14.91\%&11.65\%&16.19\%\cr
    \bottomrule
    \end{tabular}}
    \end{threeparttable}
    \label{comparisons_model}
    \vspace{-2mm}
\end{table}

\textbf{Both GNN and multi-behavior based methods improve model performance.}
Despite the various architectures among different baseline models, we can find that GNN based models have a consistent trend that perform much better than non-graph models.
For example, by incorporating neighbor information into representations, MBGCN and NGCF outperform DIPN and DMF in most datasets and metrics at the single-behavior and multi-behavior settings, respectively.
Besides, multi-behavior models KHGT and MBGCN achieve much better performance than single-behavior model KGAT and NGCF, which further verifies the effectiveness of adding multi-behavior information for learning.  
However, special designs are required to better utilize the graph information and multi-behavior information, and naive application may even drop the performance.
For example, NGCF$_M$, which uses multi-behavior information, perform much worse than MATN.

\textbf{CKML consistently outperforms GNN based multi-behavior baseline models.} 
Our proposed CKML surpasses in the performance of NGCF$_{M}$, LightGCN$_{M}$, MBGCN, and the state-of-the-art multi-behavior model KHGT. 
By empowering the multi-behavior recommendation with multi-interest learning, CKML is capable of modeling the complex dependencies among multiple behaviors with multi-grained representations to infer user preference.
While existing multi-behavior models only consider the observed user-item interactions as unified representations.

\begin{table}[H]
\setlength{\abovecaptionskip}{0cm}
\setlength{\belowcaptionskip}{-3mm}
\caption{Training time comparison (seconds per epoch) of different methods on all three datasets.}
    \centering
\setlength{\tabcolsep}{1mm}{
\small
\begin{tabular}{c|c|c|c}
\midrule[0.25ex]
\diagbox{Model}{Training time (s)}{Dataset} &
\multicolumn{1}{c|}{Yelp} &
\multicolumn{1}{c|}{MovieLens} & 
\multicolumn{1}{c}{Retail} \\ \hline 
KHGT & 46.65 & 56.11 & 74.31      \\
CKML & 40.29 & 44.07 & 59.88     \\  
\hline 
\end{tabular}}
\label{tab:time_cmp}
\vspace{-3mm}
\end{table}

\subsubsection{Efficiency Comparison}
In addition to effectiveness, efficiency is also important.
Table \ref{tab:time_cmp} shows the average training time of our proposed CKML and KHGT for each epoch. 
For the sake of fairness, we set the parameters related to training efficiency consistent, like batch size and GNN layer.
We can find that CKML is faster with \textbf{13.63\%}, \textbf{21.46\%}, and \textbf{19.42\%} time reduction on the three datasets. 
One probable reason is that we split the complete graph into several smaller graphs under interests, and then make computation separately on these smaller graphs, which can be accelerated by parallel computation. 


\begin{table}[H]
    \setlength{\abovecaptionskip}{0cm}
    \setlength{\belowcaptionskip}{-3mm}
    \caption{Performance of different CKML variants. The significance level $p$-value for all data in the table is $<0.05 $.}
    \centering
    \begin{threeparttable}
    \resizebox{\linewidth}{!}{
    \begin{tabular}{c|cccccc}
    \toprule
    \multirow{2}{*}{Model}&
    \multicolumn{2}{c}{Yelp}&\multicolumn{2}{c}{MovieLens}&\multicolumn{2}{c}{Retail}\cr
    \cmidrule(lr){2-3} \cmidrule(lr){4-5} \cmidrule(lr){6-7}
    &HR&NDCG&HR&NDCG&HR&NDCG\cr
    \midrule
    CKML &\textbf{0.896}&\textbf{0.624}&\textbf{0.927}&\textbf{0.686}&\textbf{0.527}&\textbf{0.323}\cr
    CKML w/o CIE &0.893&0.619&0.904&0.629&0.510&0.310\cr
    CKML w/o FBC &0.887&0.610&0.884&0.598&0.491&0.290\cr
    CKML w/o MI &0.839&0.524&0.856&0.580&0.444&0.246\cr

    \bottomrule
    \end{tabular}}
    \end{threeparttable}
    \vspace{-5mm}
    \label{tab:ablation}
\end{table}

\subsection{Ablation Study}

CKML is built with several important designations including the Multi-Interest (MI), the Coarse-grained Interest Extracting (CIE) and the Fine-grained Behavioral Correlation (FBC).
To analyze the rationality of each design consideration, we explore CKML with several different model variants. 
\textbf{CKML w/o CIE}: We remove the coarse-grained interest extracting module and express each interest with randomly initialized vectors. \textbf{CKML w/o FBC}: We replace the fine-grained behavioral correlation module with a combination of  the best-performing GCN methods (LigthGCN for Yelp and MovieLens, GCCF for Retail) and summation operation. \textbf{CKML w/o MI}: To evaluate the effectiveness of multi-interest, we remove the above two modules simultaneously and use unified vectors for users and items representations.
The performance of CKML and its variants are summarized in Table \ref{tab:ablation}.
We can obtain the following observations:
\begin{itemize}
\item Comparing the performance of CKML and its first three variants, we can find that each variant brings about performance degradation when any key component is removed or replaced with other modules.
This demonstrates the rationality and effectiveness of the three key designations.
\item It is worthwhile noticing that CKML w/o MI achieves the worst performance on all three datasets compared to other variants with multi-interest learning.
In particular, this variant has a performance decline up to 16.03\%, 15.45\% and 23.84\% in terms of NDCG on Yelp, MovieLens and Retail datasets.
This further demonstrates the effectiveness of multi-interest for the modeling of the complex dependencies among multiple behaviors.
\end{itemize}

\begin{table}[H]
   \setlength{\abovecaptionskip}{0cm}
    \caption{Impact of share interests and specific interests}
    \centering
    \begin{threeparttable}
    \resizebox{\linewidth}{!}{
    \begin{tabular}{c|cccccc}
    \toprule
    \multirow{2}{*}{Model}&
    \multicolumn{2}{c}{Yelp}&\multicolumn{2}{c}{MovieLens}&\multicolumn{2}{c}{Retail}\cr
    \cmidrule(lr){2-3} \cmidrule(lr){4-5} \cmidrule(lr){6-7}
    &HR&NDCG&HR&NDCG&HR&NDCG\cr
    \midrule
    CKML-Shared & 0.896&0.620&0.834&0.546&0.513&0.311\cr
    CKML-Specific&0.814&0.497&0.925&0.680&0.271&0.140\cr
    CKML &\textbf{0.896}&\textbf{0.624}&\textbf{0.927}&\textbf{0.686}&\textbf{0.527}&\textbf{0.323}\cr
    \bottomrule
    \end{tabular}}
    \end{threeparttable}
    \vspace{-0.4cm}
    \label{tab:share_specific}
\end{table}

\subsection{Study of Interests}
We propose to explicitly separate interests into shared and specific interests to alleviate the negative impact of irrelevant interactions.
To demonstrate the superiority of this correlation modeling strategy, 
we replace it with two variants, namely, only shared interests and only specific interests.
We keep the number of interests fixed and apply them as the basis of CKML for multi-behavior recommendation.
Resulted variants are named as CKML-Shared and CKML-Specific respectively.
The results are reported in Table \ref{tab:share_specific}. There are some observations:
\begin{itemize}
    \item We can find that CKML-Shared achieves poor performance on MovieLens dataset.
    Possible reason is that the behavior types in this dataset are Like, Neutral and Dislike, which are mutually exclusive with each other.
    So CKML-Shared may suffers from the negative impact of irrelevant interactions from other behaviors on MovieLens dataset. 
    Besides, we also find that there exists no intersection of interaction data between different behaviors on MovieLens, that is, a user can interact with the same item through at most one type of behavior.
    This further validates our assumption.
    \item CKML-Specific performs worse on Yelp and Retail datasets. 
    This is because CKML-Specific fails to utilize information of other behaviors to assist the recommendation of target behavior as it neglects shared interests among multiple behaviors (e.g., Tip and Like on Yelp, as well as Add-to-cart and Purchase on Retail).
    \item CKML, which considers shared and specific interests, achieves the best performance on all three datasets. 
    It suggests that taking into account both share and specific interests eliminate the effect of irrelevant interactions and improve the robustness of the model.
\end{itemize}

\begin{figure}[t]
	\centering
	\setlength{\belowcaptionskip}{0cm}
	\setlength{\abovecaptionskip}{0cm}
	\includegraphics[width=0.47\textwidth]{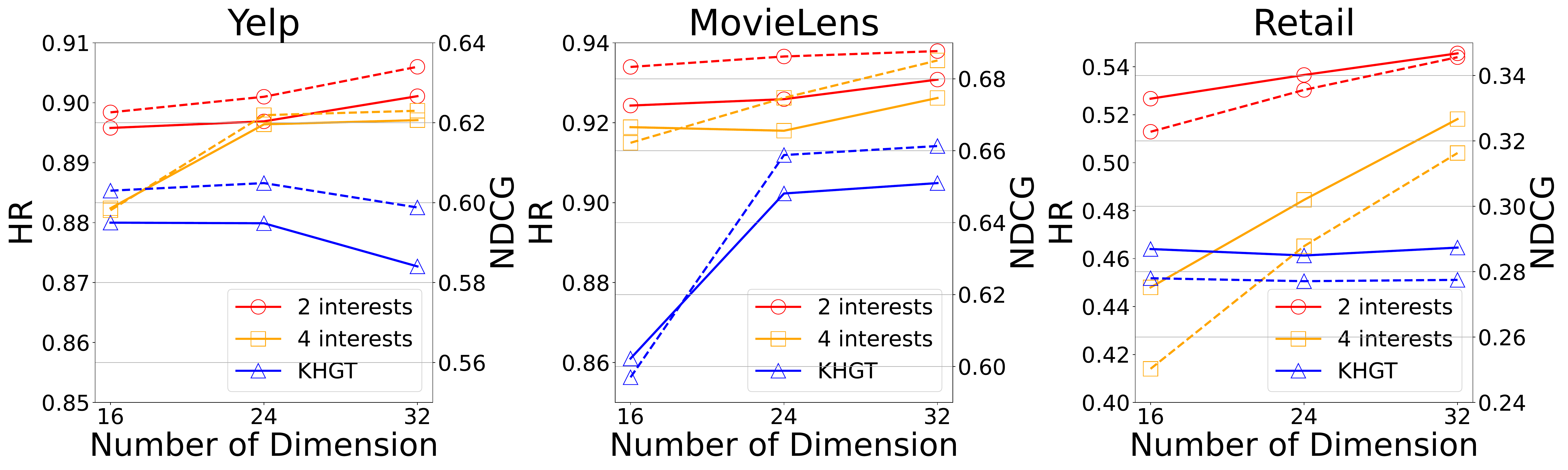}
    \caption{Impact of the number of interests}
	\label{fig:num_interest}
	\vspace{-3mm}
\end{figure}

\begin{figure}[!t]
	\centering
	\setlength{\belowcaptionskip}{0.1cm}
	\setlength{\abovecaptionskip}{0.1cm}
	\includegraphics[width=0.47\textwidth]{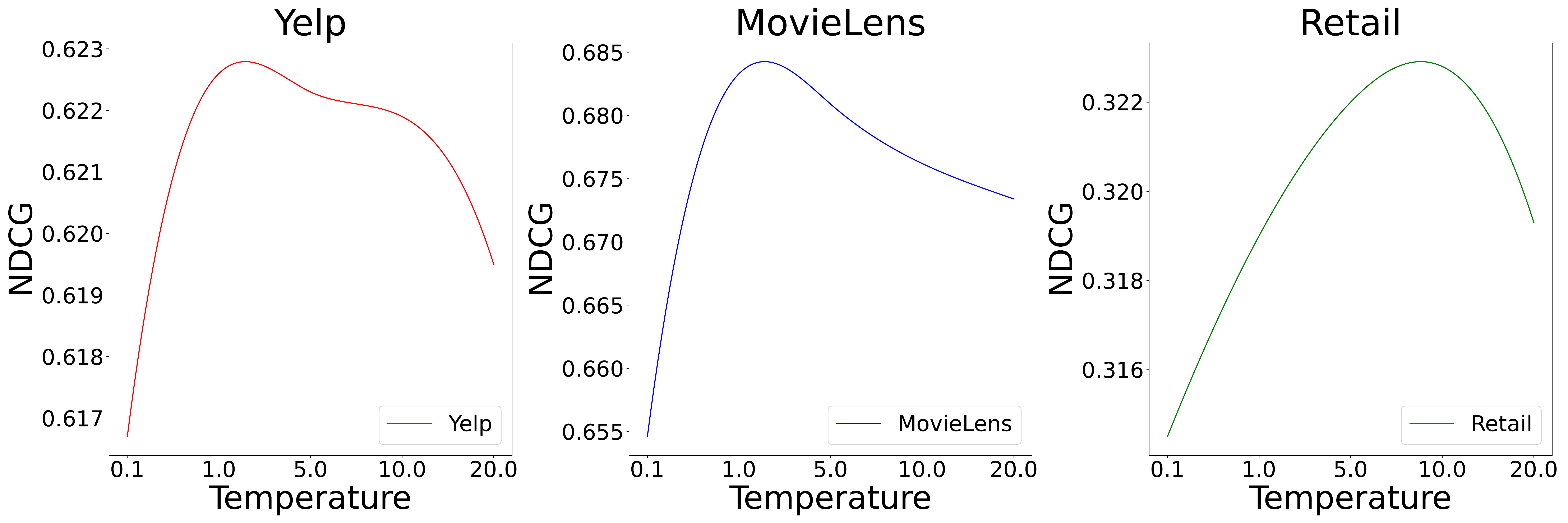}
    \caption{Impact of temperature coefficient}
	\label{fig:temperature}
	\vspace{-3mm}
\end{figure}

\subsection{Hyper-parameter study}
\subsubsection{Impact of the number of interests}
\label{section:num_interest}
To investigate how the number of interests affects the performance of CKML, we adjust the number of interests in the range \{2,4\}. 
For simplicity, we set the number of shared interests and specific interests to the same. 
The results are presented in Figure \ref{fig:num_interest}.
We can find that when embedding size is set to 16 in line with KHGT, the model with 2 interests achieved the best results on all three datasets. Performance drops a lot when the number of interests increases from 2 to 4. 
Possible reason may be the too small embedding size (only 8) of each interest which can hardly learn good representations.
We further extend the embedding size to 16 and 32, and we can observe that our model achieves significant performance improvement for both 2 interests and 4 interests.
This verifies our above assumption.
 When embedding size grows larger, KHGT performs consistently worse than our model, which shows the superior performance of our proposed CKML.
Moreover, KHGT has a performance drop on Yelp and Retail dataset when a larger embedding size is applied.
Possible reason is that KHGT is easier to overfit due to the overlooking of multi-interest.

\subsubsection{Impact of temperature coefficient}
\label{section:temperature}
The interactions between users and items are due to a single interest or the combination of multiple interests.
To investigate it, we change the temperature coefficient used for behavior allocation and the results are reported in Figure \ref{fig:temperature}. 
We can see that a moderate temperature coefficient is needed for the CKML to achieve the best performance.
And when the temperature coefficient is set too small, the performance of the model deteriorates rapidly.
One possible reason is that the probability distribution of interest is close to the one-hot vector in this case, which makes it challenging to learn. 
Besides, the model performance degrades either if the temperature coefficient is set too large. 
This may be because the weights of multiple interests become similar, and the model fails to identify the interest behind the interaction well. 
This again illustrates the importance of exploring multiple interests.

\begin{figure}[t]
	\setlength{\belowcaptionskip}{-0.3cm}
	\setlength{\abovecaptionskip}{0cm}
	\subfigure{
        \begin{minipage}[t]{0.48\linewidth}
        \centering
		\label{fig:agg_hr} 
		\includegraphics[width=1.6in]{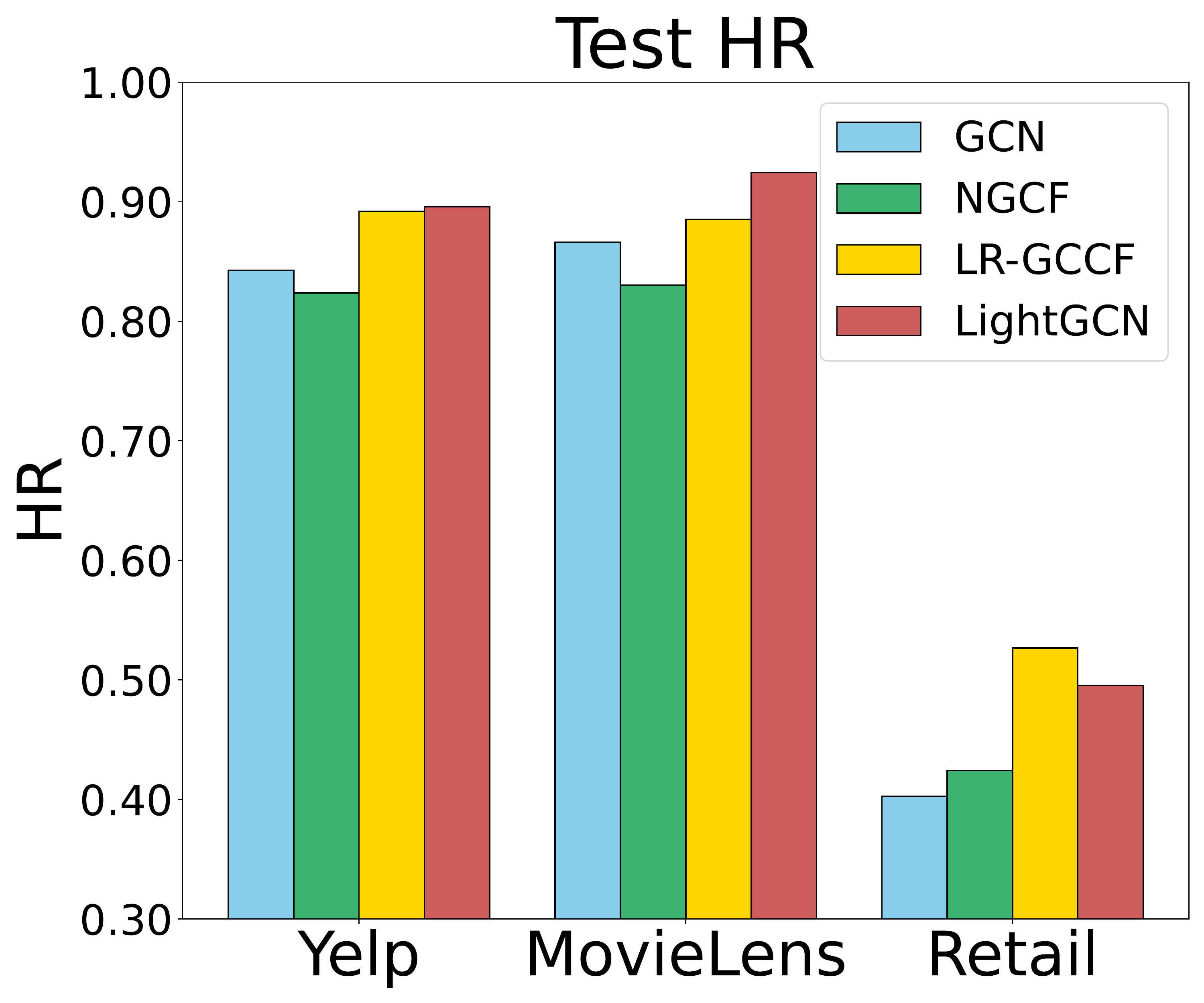}
        \end{minipage}}
	\subfigure{
        \begin{minipage}[t]{0.48\linewidth}
        \centering
		\label{fig:agg_ndcg} 
		\includegraphics[width=1.6in]{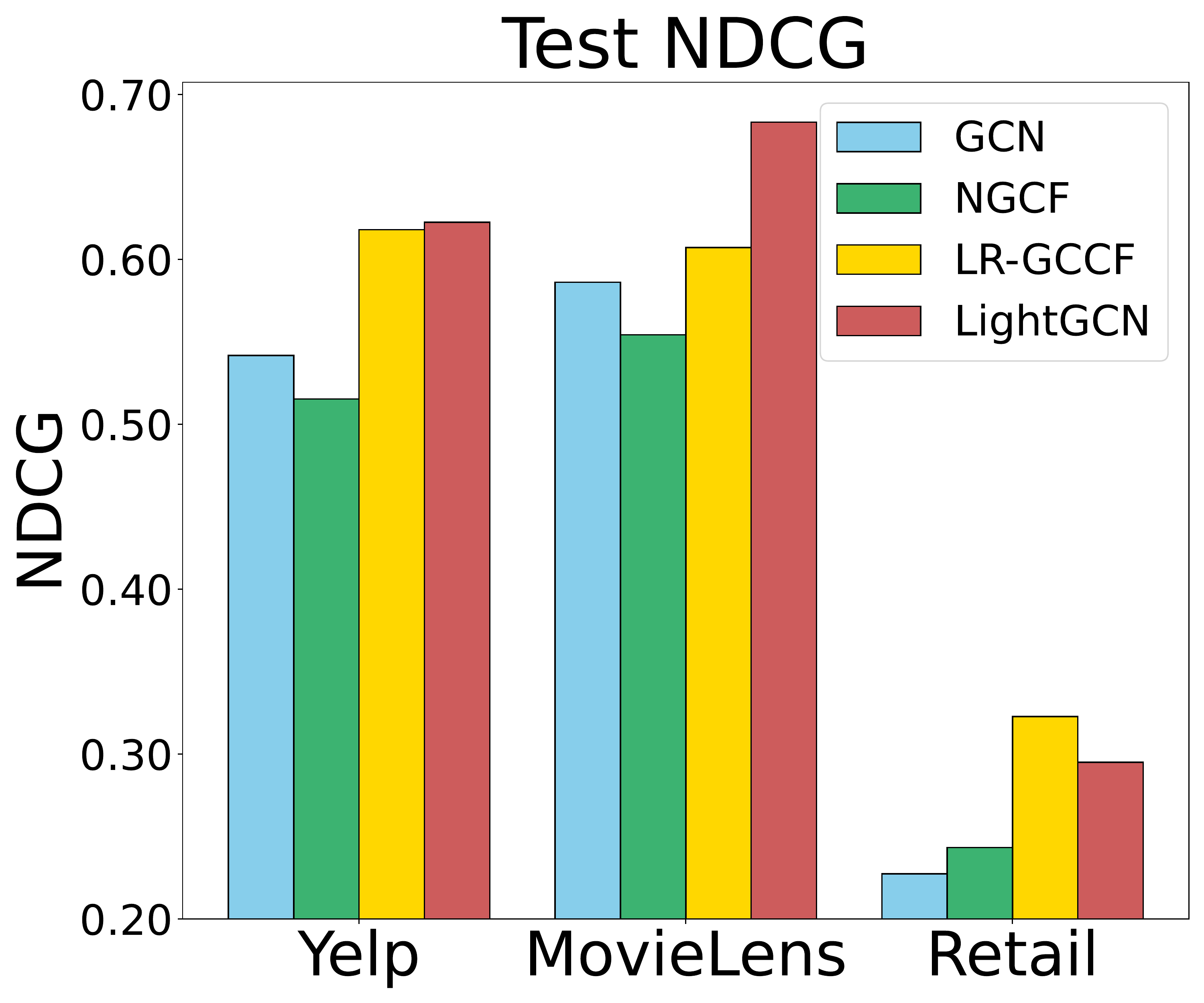}
        \end{minipage}}
	\caption{Impact of GCN aggregators.}
	\label{fig:figureaggregators}
\end{figure}

\subsubsection{Impact of GNN aggregators.}\label{Imapct_of_aggregatoe}
We investigate the impact of different GNN aggregators i.e. GCN \cite{gcn}, NGCF \cite{ngcf}, LR-GCCF \cite{lr-gccf} and LightGCN \cite{lightgcn}. 
The models with different aggregators are compared in Figure \ref{fig:figureaggregators}. 
We can find that LightGCN performs the best on Yelp and MovieLens among the four aggregators. 
The reason might be that removing the transformation matrix and nonlinear functions enables easier training and alleviate overfitting. 
CKML with LR-GCCF achieves the best performance on Retail, probably because Retail contains multiple types of closely correlated behaviors, which has high requirements on the fitting ability of the model. So the introduction of a nonlinear activation function better facilitates the model to fit Retail.

\subsection{Case Study}

\begin{figure}[!t]
	\centering
	\setlength{\belowcaptionskip}{0.1cm}
	\setlength{\abovecaptionskip}{0.1cm}
	\includegraphics[width=0.47\textwidth]{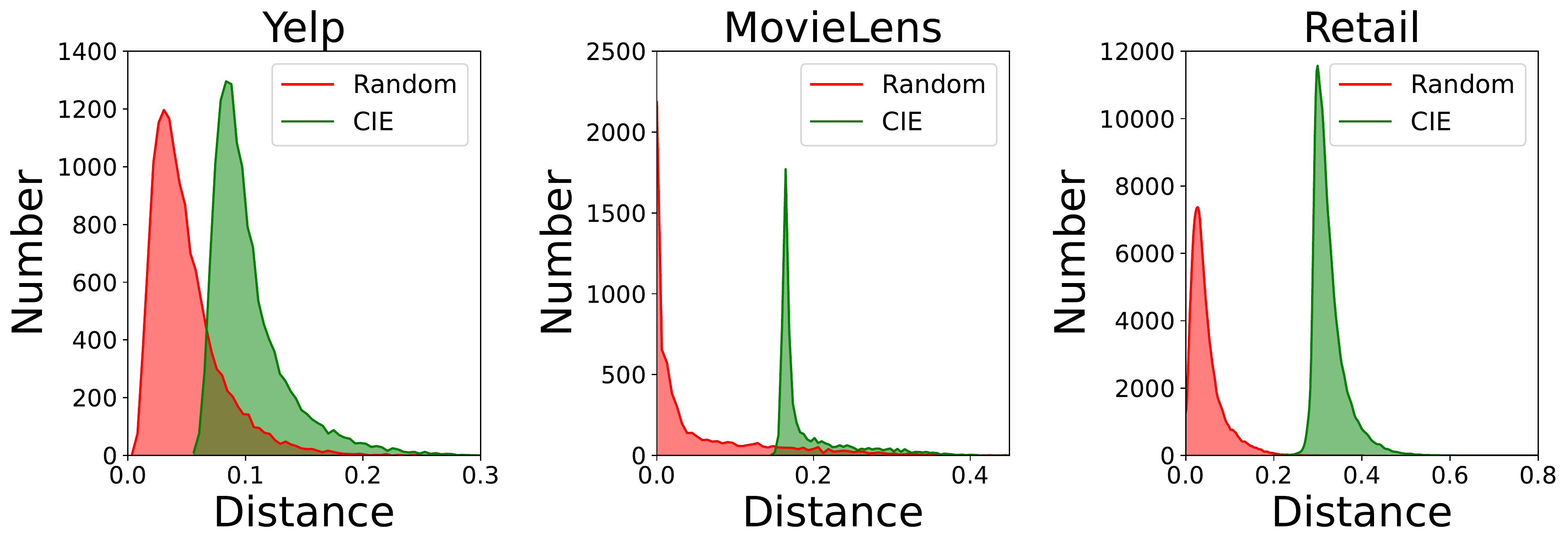}
    \caption{The distribution of average distances.}
	\label{fig:distance_shared_specific}
	\vspace{-5mm}
\end{figure}

\subsubsection{The visualized analysis of interest initialization.}
We have claimed in Section \ref{intro} that initializing the clustering centers to be far apart is significant for the learning of interests. 
To validate that the initialized representations of CIE are better than Random, we calculate the average Euclidean distance among different interests on all items for CIE and random, then plot the distance distribution in Figure \ref{fig:distance_shared_specific}.
We can observe the overall distribution of the average distances of interests obtained by CIE is further across all three datasets, which means the clustering centers initialized by CIE are farther apart than Random.
It suggests that CIE can better initialize interest centers, enabling the model to identify the interest behind interactions efficiently.

\begin{figure}[!t]
	\centering
	\setlength{\belowcaptionskip}{0cm}
	\setlength{\abovecaptionskip}{0cm}
	\includegraphics[width=0.47\textwidth]{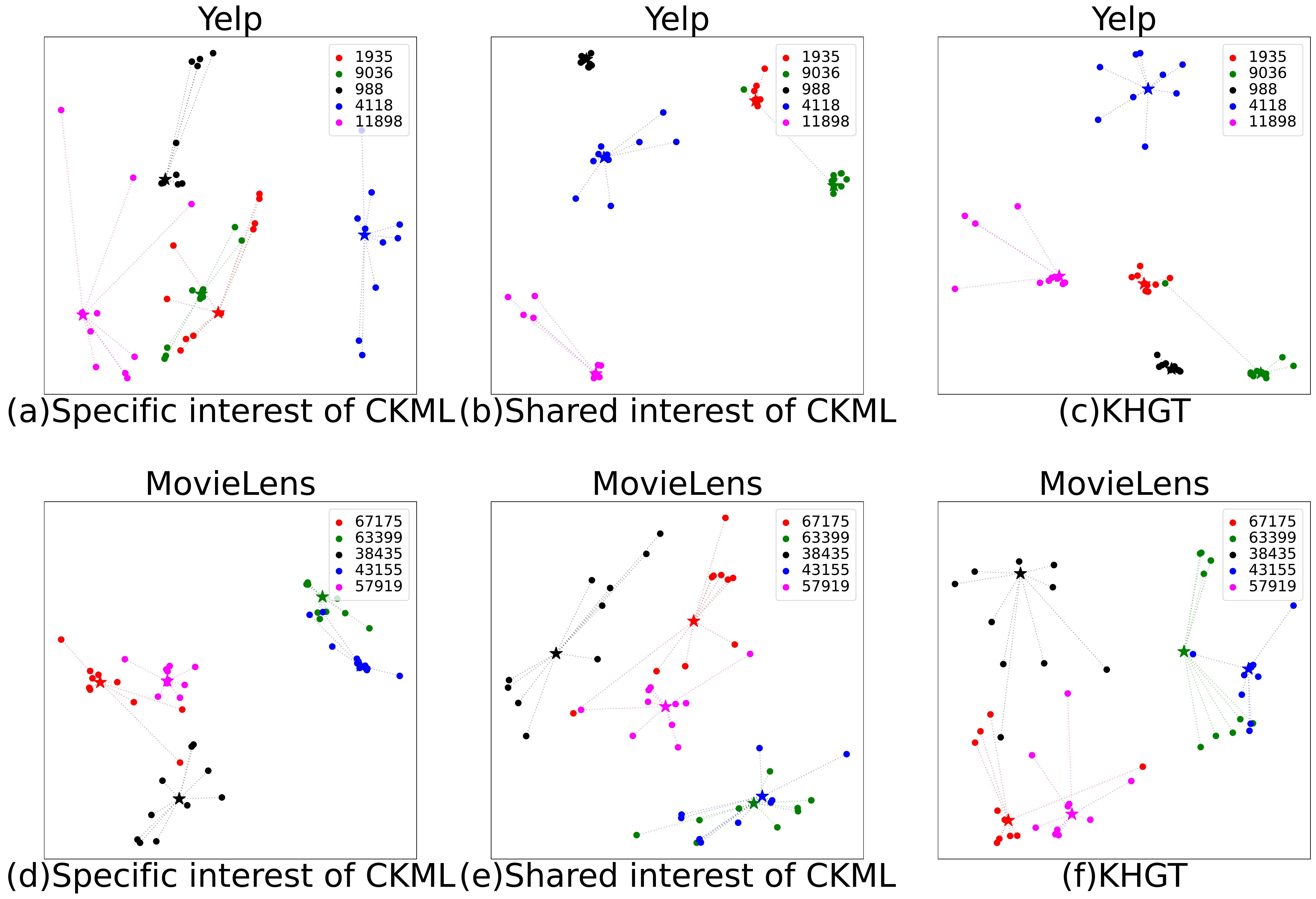}
    \caption{Visualization of items representations via t-SNE. Points of the same color represent items being interacted with by the same user. Each star is the center of points with the same color.}
	\label{fig:case_study}
	\vspace{-3mm}
\end{figure}

\subsubsection{The visualization analysis of shared and specific interests}
We randomly select five users and the items they have interacted with under the target behavior. 
In Figure \ref{fig:case_study}, we visualize the representations of items under shared interest and specific interest obtained from CKML, as well as the representations obtained by KHGT. 
 Comparing the points with the same color in Figure \ref{fig:case_study}(a)(b)(c), we can find that items under the shared interest and KHGT are more clustered than specific interest representations. A probable reason is that Yelp has few interactions of target behavior which makes it hard to mine the interest-related information behind the interaction. Besides, the shared interest and KHGT introduce additional interaction information of other behaviors, which makes it better to learn the representations of items.
 By comparing Figure \ref{fig:case_study}(d)(e)(f), we can observe that the result of specific interest of CKML exhibit much tighter clusters, showing a significant grouping effect for items that have been purchased by users. While the clustering effect of the other two is less pronounced. It is possibly due to the weak correlation between the three behaviors (e.g., like, neutral and dislike) on MovieLens dataset, which brings noise to the representation of shared interests. In all, the results illustrate the effectiveness of introducing specific interest on datasets which is similar to MovieLens.


\section{Conclusion}
\label{future works}
In this paper, we propose the CKML framework for multi-behavior recommendations. 
In order to make full use of knowledge-aware information to extract shared and behavior-specific interest representations, we propose the CIE module. 
To further learn the interest representation of each user and item under different behaviors and exchange information under different behaviors at fine granularity, we propose a GNN-based FBC module, which allocates edge weights by dynamic routing and exchanges information by self-attention mechanism.
We conduct comprehensive experiments on three real-world datasets and show that the proposed CKML outperforms all state-of-the-art methods on all three datasets.
Besides, the additional visualization experiment demonstrates the superiority of our well-designed shared and behavior-specific interests.


\begin{acks}
We thank MindSpore \cite{mindspore} for the partial support of this work, which
is a new deep learning computing framework.
\end{acks}


\newpage
\bibliographystyle{ACM-Reference-Format}
\bibliography{sample-base}

\appendix

\end{document}